\documentclass[aps,preprint,prb]{revtex4-1}
\usepackage{graphicx}

\usepackage{amsmath}
\usepackage{bm}

\usepackage{epstopdf}

\usepackage{dcolumn}
\usepackage{pstricks,pst-plot,epsfig}
\usepackage{varioref}
\usepackage{amssymb}
\usepackage{amsthm}
\usepackage{hyperref}
\usepackage{color,soul}

\usepackage[format=plain,justification=justified,singlelinecheck=false,font={stretch=1.125,small,sf},labelfont=bf,labelsep=space]{caption}
\usepackage{setspace}

\begin{document}

\title{Interaction of oxygen with pristine and defective $\rm MoS_2$ monolayers}

\author{Murilo Kendjy Onita}
\author{Flavio Bento de Oliveira}
\author{Andr\'eia Luisa da Rosa}
\email[]{andreialuisa@ufg.br}
\affiliation{Instituto de F\'isica, Universidade Federal de Goi\'as, Campus Samambaia, 74690-900, Goi\^ania, Goi\'as, Brazil.}

\date{\today}

\begin{abstract}

Atom controlled sub-nanometer MoS$_2$ pores have been recently
fabricated.  Oxidative environments are of particular interest for
MoS$_2$ applications in electronics, sensing and energy storage. In
this work we carried out first-principles calculations of oxygen
adsorption in plain and sub-nanometer MoS$_2$ nanopores.  The chemical
stability of the layers and pores towards oxygen was verified using
density-functional theory. Dissociation and diffusion barriers have
been calculated in order to understand surface and pore oxidation and its electronic
properties at the atomic scale.

\end{abstract}


\maketitle

\section{Introduction}

Owing to their fascinating properties, two dimensional transition
metal dichalcogenides (TMDs) have been explored for a variety of
applications, including electronics and optoelectronics,
photonics, catalysis and energy storage\,\cite{Feng2015,Feng2016,Nature2016,Karmodak2021,Bhim2021}.
In particular, molybdenum disulfide (MoS$_2$), the most promising TMDCs,
is efficiently exfoliated in monolayer or multilayers\cite{Santosh2015,NatComm2017}.
Recently the fabrication of MoS2 sub-nanometer pores
offer several opportunities being promising candidates for several
technological applications such as membranes for DNA
translocation,\cite{Feng2015,Sen2021,Graf2019}, water filtration and
desalination\,\cite{Cao2020,Kou2016,Nature2015,Wang2021}, energy
harvesting\,\cite{Graf2019a} and hydrogen evolution
reaction\,\cite{Wu2019,Li2019,Frenkeldefects2022}.

Control at atomic scale nanopores can be achieved by using electron or
ion beams to fabricate nanopores of specific
sizes\,\cite{SciRep2019,Shi2017,Garoli2019,NatNano2014,SciRep2016}.
Sulfur and molibdenum vacancies, the most abundant defect species in MoS$_2$, may
serve as the nucleation sites for the nanopore
formation\,\cite{NatComm2017,Trainer2022}.  Pore sizes down to 0.5-1.2\,nm 
have been created, corresponding to a few atoms
missing\,\cite{Zhao2018,Nanoscale2017,ACSNano2020,ACSAPPLIED2018}.

In particular, the
interaction of small molecules  with MoS$_2$ is of particular importance, since
it plays a role in the properties and performance of
two-dimensional-based devices. Concurrently, structural defects such as vacancies and edges are particularly susceptible to
attacks in reactive environments\,\cite{ACSNano2022}. Adsorption energy and diffusion of oxygen on MoS$_2$ is controversial and not completely
understood. Moreover, oxygen on surfaces either forms undesired alloys
entering a sulphur site or possesses a low binding energy. One important, not completely understood question, is the role of point defects on the adsorption of oxygen on MoS2. Previous experimental investigations suggested that the presence of defects significantly alters the
surface stability\,\cite{Santosh2015,Longo2017}. Oxygen leads to the formation of Mo oxide in the two-dimensional
MoS$_2$ lattice, yielding an overall disordered and fragmented
structure. On the other hand, other investigations suggested that oxygen  is incorporated in the MoS$_2$ lattice at substitutional sites\,\cite{Jeon2015}.
X-ray photoelectron spectroscopy measurements have shown that the basal plane of
MoS$_2$ monolayers, when subjected to long-term ambient exposure,
spontaneously undergoes oxygen substitution reactions, giving rise to
a highly crystalline two-dimensional molybdenum oxysulfide
phase\,\cite{Peto2018,NatChem2013}.

In this work we have investigated the interaction of oxygen atoms and
molecules with MoS$_2$ monolayers and subnanometer pores. The pore
size and termination  play a crucial role on the
interaction with oxygen molecules.

\section{Methodology}

All calculation have been performed using the density functional
theory \cite{Hohenberg:64,Kohn:65} as implemented in the
VASP\,\cite{Kresse:99}. For the exchange-correlation functional we use
the GGA approximation\,\cite{Perdew:96}.  Basis set consisting of an
expansion in plane waves and an energy cutoff of 300\,eV were used.
The Brillouin zone was sampled according to the Monkhorst-Pack
scheme\,\cite{MP} with $\Gamma$ point only in a $(6\times6)$ unit
cells. To avoid interaction between neighbouring wires, a vacuum space
of 15\,{\AA} was set in the direction perpendicular to the
two-dimensional MoS$_2$ layers. The structural relaxation was
performed until the atomic forces were less than
10$^{-3}$\,eV/{\AA}. In order to determine the diffusion and reaction
barriers we have performed CI-NEB\cite{NEB1,NEB2} calculations with at least seven
images to search the minimum-energy reaction paths and saddle points
between the initial state and final state configurations.


\section{Results}

\begin{figure}[h]
\begin{center}
\includegraphics[width=7.0cm, keepaspectratio, clip= true]{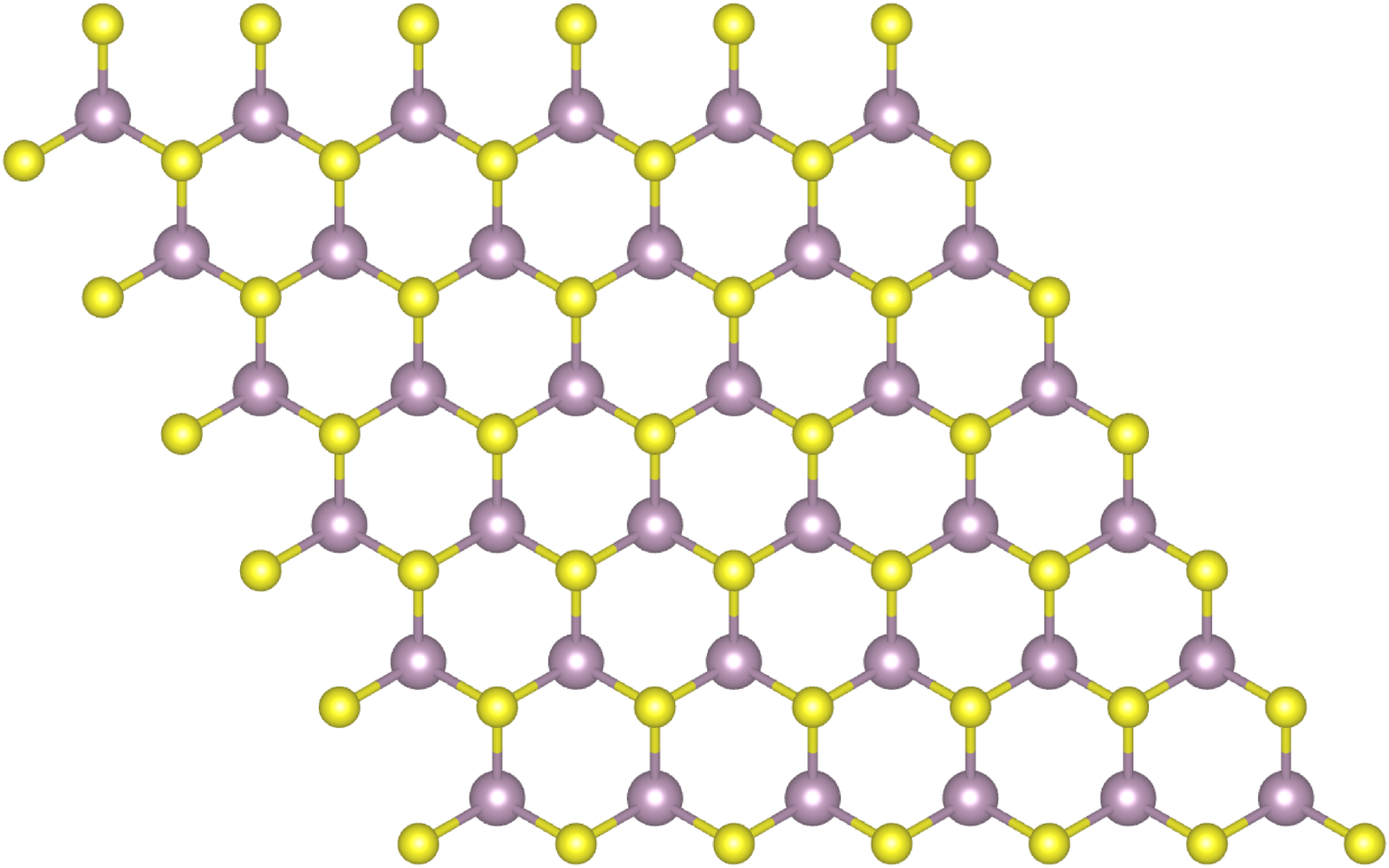}
\includegraphics[width=6.0cm, keepaspectratio, clip= true]{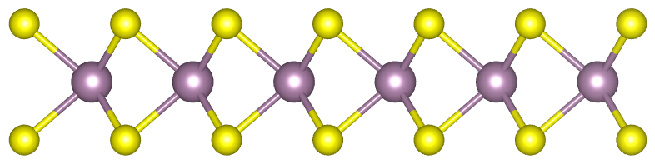}
 \includegraphics[width=8.0cm,clip = true]{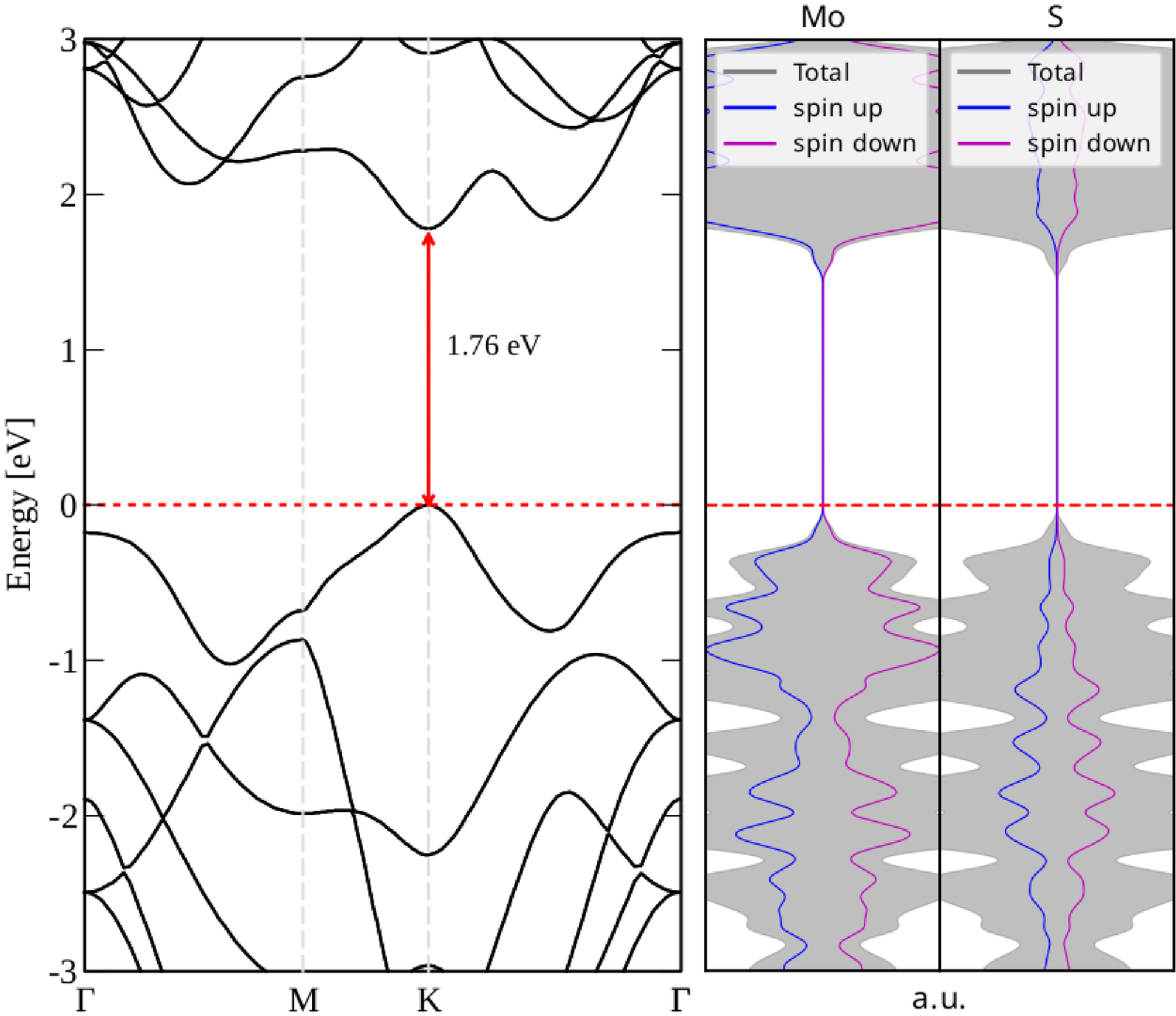}
  \caption{\label{fig:baredos} a) Top view, b) side view and c) electronic band structure and DOS of bare MoS$_2$ monolayers.}
\end{center}
  \end{figure}

Among the various arrangements for MoS$_2$, the energetically most stable one at
room temperature is known as 2H which has a honeycomb structure as
shown in Fig.\,\ref{fig:baredos} (a) and (b).  This arrangement has in
its unit cell one molybdenum and two sulfur atoms. The optimized Mo-S
(S-S) distances are 2.42 (3.14)\,{\AA}. The lattice parameter
$a$ = $b$ equals to 3.16\,{\AA}. The electronic band structure and
total density-of-states (DOS) of bare a MoS$_2$ monolayer are shown in
Fig.\,\ref{fig:baredos}(b). The direct band gap at K--point is 1.76\,eV, in
agreement with previous results\,\cite{Krasheninnikov}.

The electronic band gap and formation energies of bare and defective
monolayered MoS$_2$ are reported in our previous
publication\,\cite{JPCM2022}. The formation enthalpy of a bare MoS$_2$
monolayer is -2.63\,eV. 

Here the
adsorption energy of oxygen is calculated at high oxygen chemical
potential according to: ${\rm E_{b}= E_{MoS_2/O} - E_{MoS_2-bare} -
  \sum_i{\mu_O}}$, where ${\rm E_{MoS_2/O}}$ is the total e\ nergy of
the MoS$_2$ the with oxygen adsorbed, ${\rm E_{MoS_2}^{bare}}$ is the
total energy of the bare sheet, $\mu_{\rm O}$ is the chemical potential of
oxygen, which has ben chosen as the total energy of the oxygen
molecule. 

\begin{figure*}[ht]
\begin{center}
\includegraphics[width=6.0cm,clip = true]{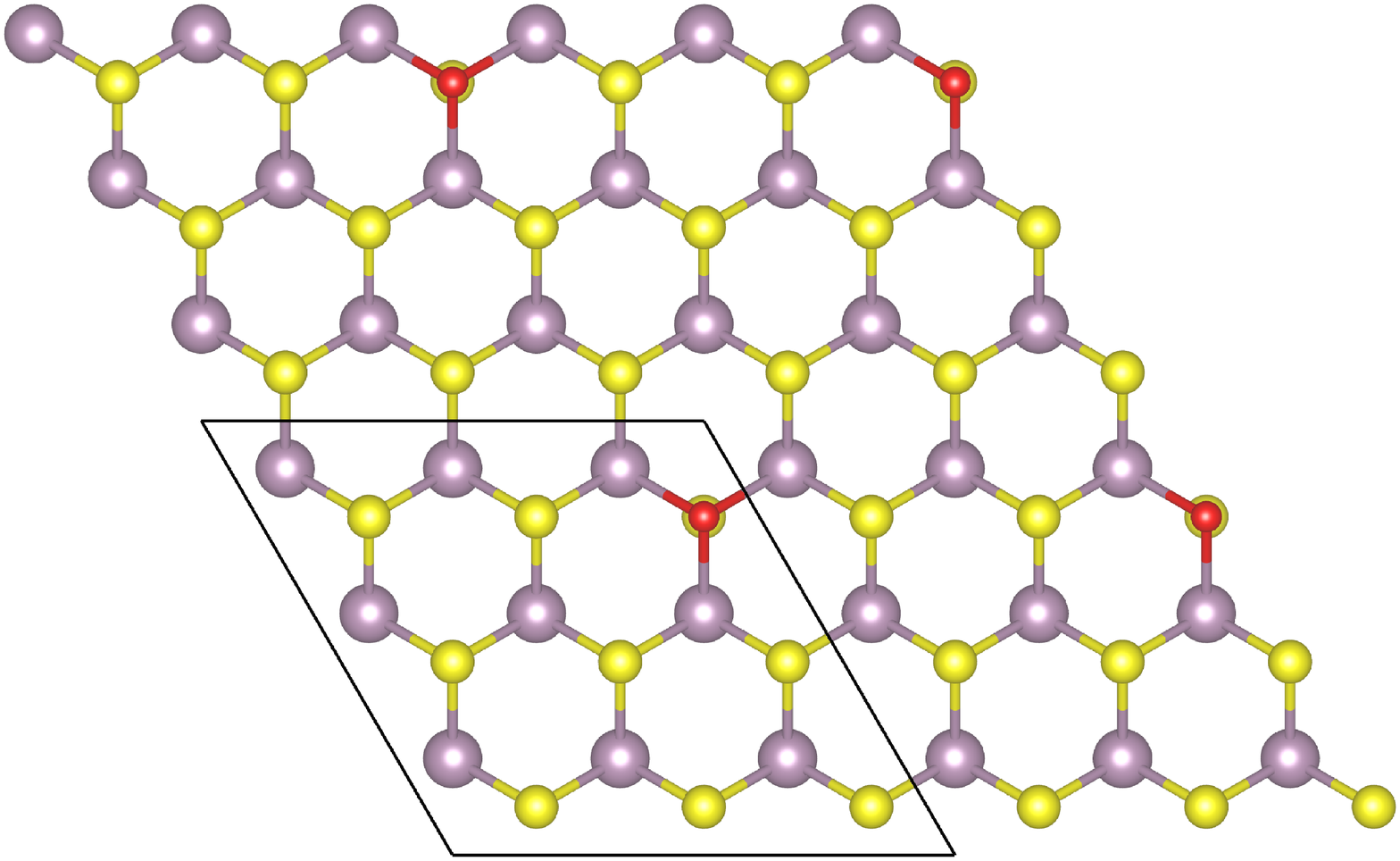}
\hspace{1cm}
\includegraphics[width=6.0cm,clip = true]{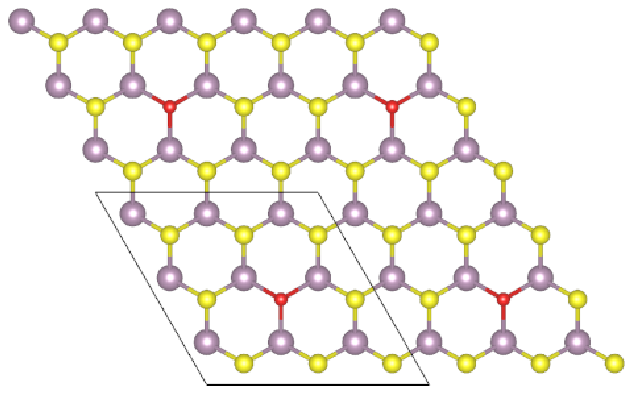}
\vspace{0.5cm}
\includegraphics[width=6.0cm,clip = true]{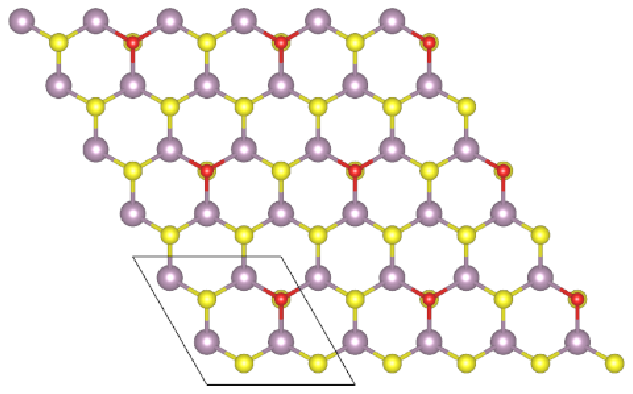}
\hspace{1cm}
\includegraphics[width=6.0cm,clip = true]{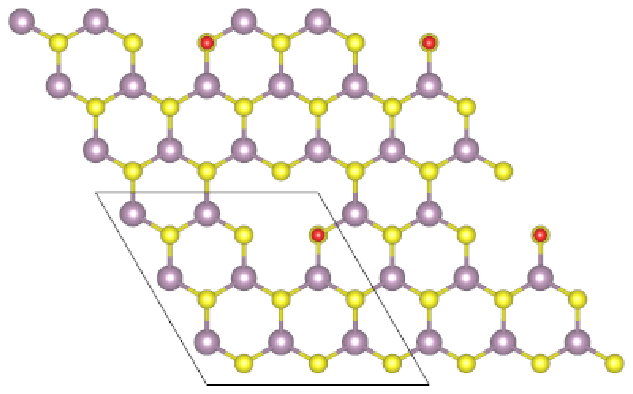}
\vspace{0.5cm}
\includegraphics[width=6.0cm,clip = true]{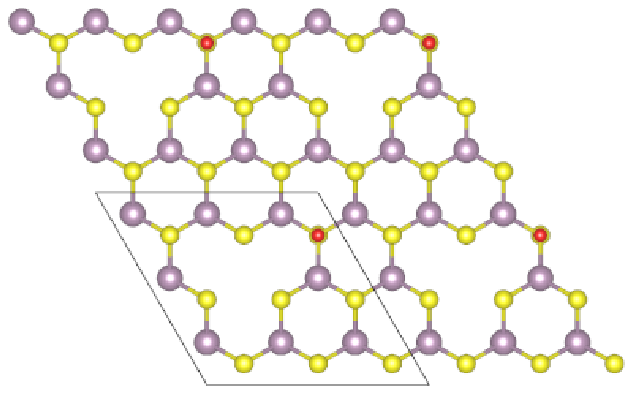}
\hspace{1cm}
\includegraphics[width=6.0cm,clip = true]{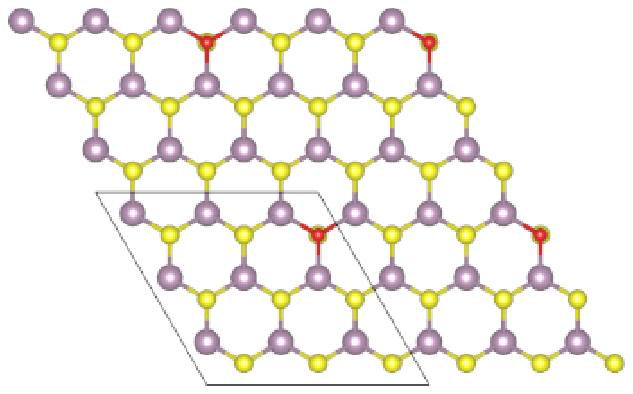}
\caption{\label{fig:oxygensub} Relaxed sructure of oxygen substitutional in ${\rm MoS_2}$ nanopores. a) O$_{\rm S}$ (11\%), b) O$_{\rm S}$ (22\%),
  c) O$_{\rm S}$ (25\%), d) O$_{\rm S}$ (11\%) + V$_{\rm Mo}$, e) O$_{\rm S}$ (11\%) + V$_{\rm S}$ (far) and f) O$_{\rm S}$ (11\%) + V$_{\rm S}$ (close). Purple, yellow and red are Mo, S and O atoms, respectively. The oxygen concentration is given within brackets.}
\end{center}
\end{figure*}

As possible configurations for defective
MoS$_2$, we have considered oxygen substituting sulphur atoms, namely O$_{\rm S}$, with concentrations of 11\%, 22\% and 25\% as shown in
Figs.\ref{fig:oxygensub} (a), (b) and (c), respectively. We notice that substitution at Mo sites is highly unfavorable. Furthermore,
defects have been incorporated considering Mo vacancies namely O$_{\rm S}$ (11\%) + V$_{\rm Mo}$, Fig.\ref{fig:oxygensub} (d) and S vacancies,
Figs.\,\ref{fig:oxygensub} (e) and  (f) labeled as O$_{\rm S}$ (11\%) + V$_{\rm S}$ (far) and O$_{\rm S}$ (11\%) + V$_{\rm S}$  (close). These concentrations were chosen
based on experimental results reported in Ref.\,\cite{Small2020}. We
have included disorder effects by adding Mo vacancies far and close to
oxygen substitutional atoms, as mentioned above.

\begin{figure*}[h!]
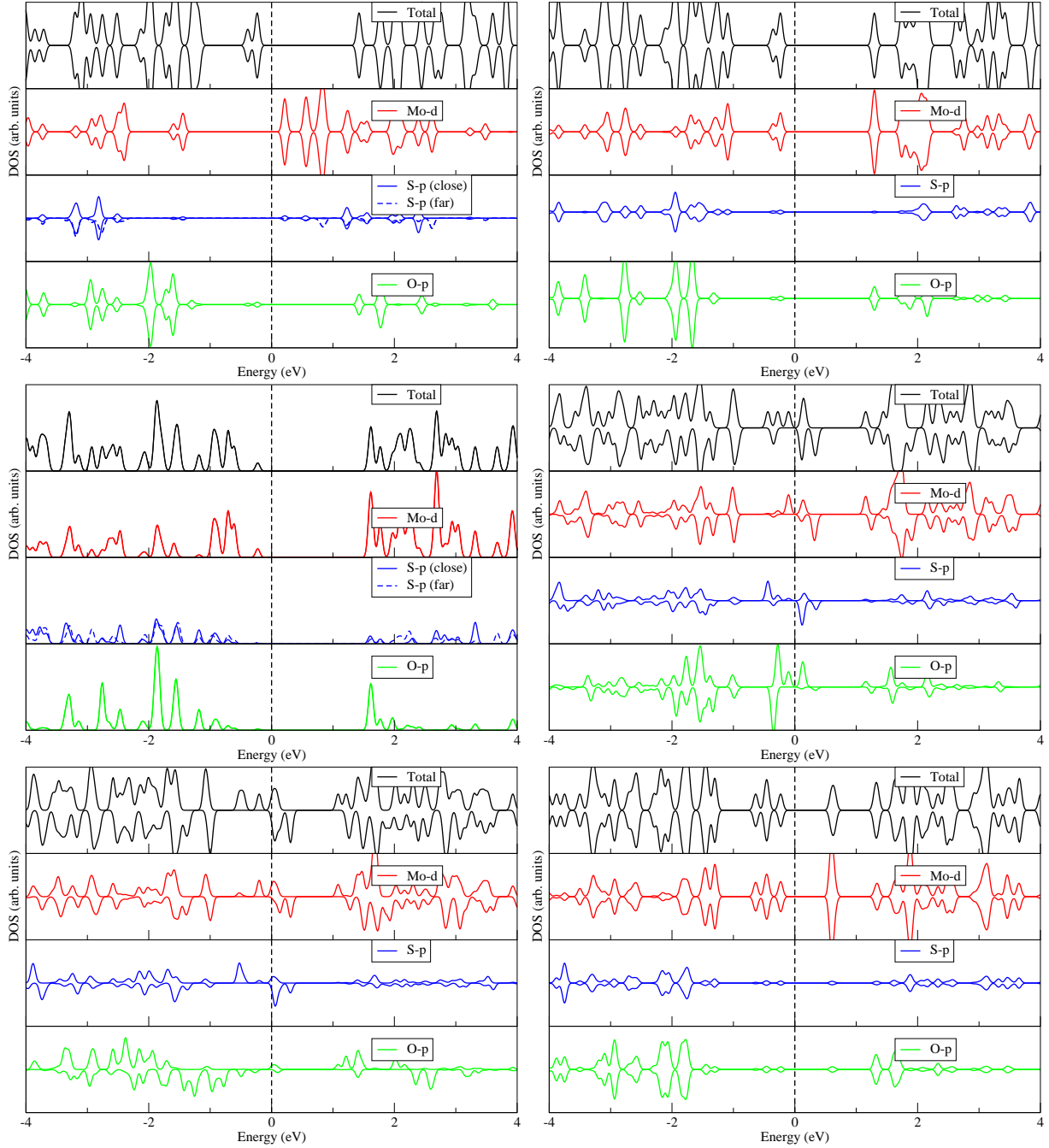

\begin{center}
\includegraphics[width=8.0cm, clip= true]{dos_MoS2_O_11.eps}
\includegraphics[width=8.0cm,clip = true]{dos_MoS2_O_22.eps}\\
\includegraphics[width=8.0cm,clip = true]{dos_MoS2_O_25.eps}
\includegraphics[width=8.0cm,clip = true]{dos_MoS2_O_11_Mo_vac.eps}\\
\includegraphics[width=8.0cm,clip = true]{dos_MoS2_O_11_Mo_vac_far.eps}
\includegraphics[width=8.0cm,clip = true]{dos_MoS2_O_11_S_vac.eps}
\caption{\label{fig:dos_oxygen_sub} DOS and PDOS of O$_S$ in $\rm MoS_2$ nanosheets at different oxygen concentrations:  a) O$_{\rm S}$ at (11\%), b) O$_{\rm S}$ at (22\%), c)  O$_{\rm S}$ (25\%), d) O$_{\rm S}^{\rm close}$ + V$_{\rm Mo}$ (11\%), e)  O$_{\rm S}^{\rm far}$ + V$_{\rm Mo}$ (11\%), f) O$_{\rm S}^{\rm close}$ +  V$_{\rm S}$ (11\%).}
\end{center}
\end{figure*}

In Fig.\,\ref{fig:dos_oxygen_sub} the DOS and PDOS of the oxygen substitutional in MoS$_2$ is shown. a) O$_{\rm S}$ at (11\%), b) O$_{\rm S}$ at (22\%), c) O$_{\rm S}$ (25\%), d) O$_{\rm S}^{\rm close}$ + V$_{\rm Mo}$ (11\%), e) O$_{\rm S}^{\rm far}$ + V$_{\rm Mo}$ (11\%), f) O$_{\rm S}^{\rm close}$ + V$_{\rm S}$ (11\%). The electronic structure of a) O$_{\rm S}$ at (11\%), b) O$_{\rm S}$ at (22\%) is very similar. The later two systems exhibit semiconductor behavior.

At higher concentration as in O$_{\rm S}$ (25\%), Fig.\ref{fig:dos_oxygen_sub}(c), additional state in the gap appear, but the structure remain semiconducting.
For  d) O$_{\rm S}^{\rm close}$ + V$_{\rm Mo}$ (11\%) and O$_{\rm S}^{\rm far}$ + V$_{\rm Mo}$ (11\%) states cross the Fermi level, indicating a metalic behavior.
Finally, for  O$_{\rm S}^{\rm close}$ + V$_{\rm S}$ (11\%) shown in Fig.\,\ref{fig:dos_oxygen_sub}(f) the system is a semiconductor as well. 
 
\begin{table}[ht!]
\caption{Binding energies of oxygen molecules in ${\rm MoS_2}$ nanopores.}
\begin{tabular}{lcc}
\hline 
\hline
Structure                             &  E$_{\rm b}$ (eV) \\
\hline
pristine                             & -0.55         \\ 
V$_{\rm 1Mo-O_2}$ (non-diss.)         &  -0.37       \\
V$_{\rm 1S-O_2}$  (diss.)             &  -0.90        \\
V$_{\rm 2S-O_2}$ (diss.)              & -3.44        \\ 
V$_{\rm 1Mo6S-3O_2}$(diss.)          &  -3.24     \\
V$_{\rm 1Mo6S-O_2}$ (non-diss.)       &   -0.45      \\ 
\hline
\hline
\end{tabular}
\label{tab:adsorption}
\end{table}

\begin{figure*}[h!]
\begin{center}
{\includegraphics[width=4.0cm,clip = true, keepaspectratio]{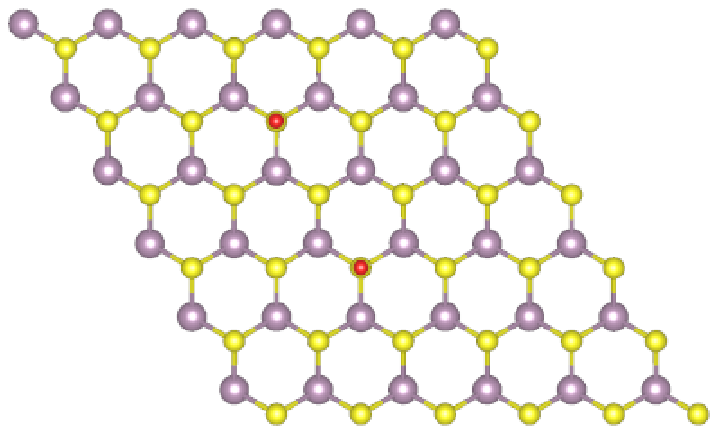}}
\vspace{0.5cm}
\includegraphics[width=4.0cm,  keepaspectratio, clip = true]{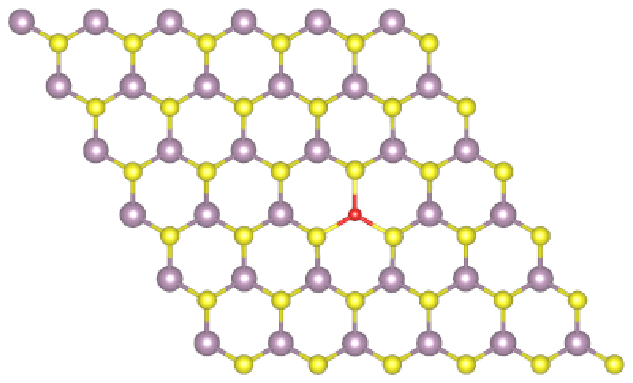}
\vspace{0.5cm}
\includegraphics[width=4cm,  keepaspectratio, clip = true]{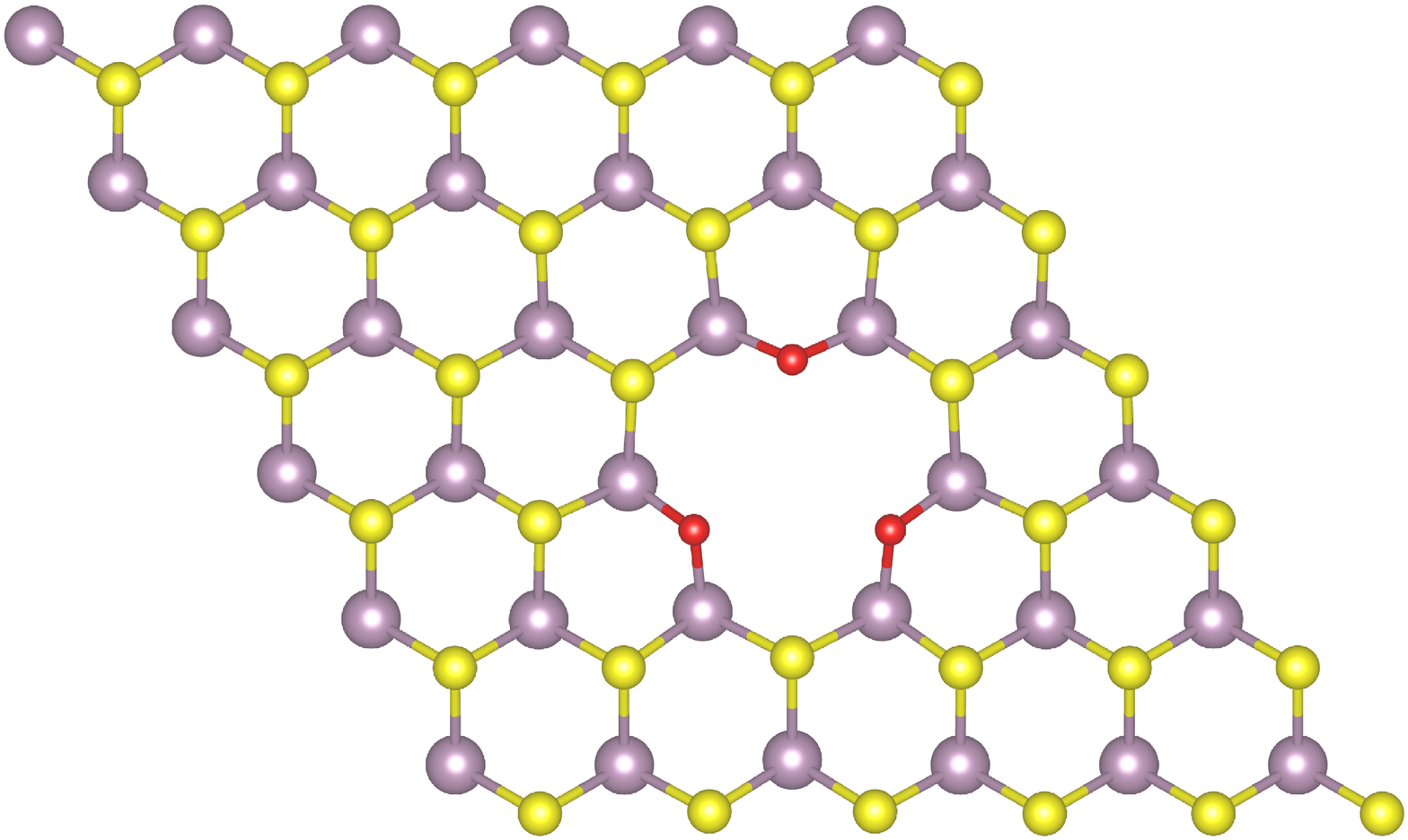}
\vspace{0.5cm}
\includegraphics[width=4.0cm,  keepaspectratio, clip = true]{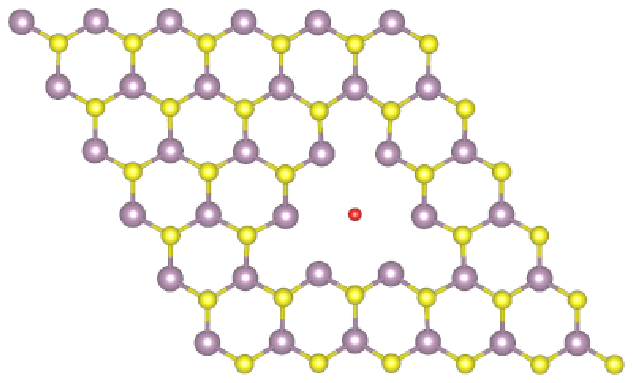}\\
\includegraphics[width=3cm,clip = true, keepaspectratio]{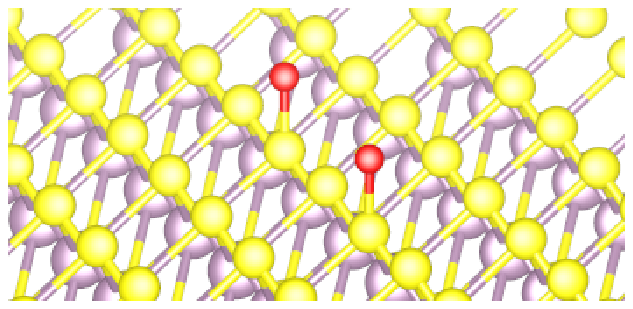}
\hspace{0.5cm}
\includegraphics[width=3cm, keepaspectratio, clip = true]{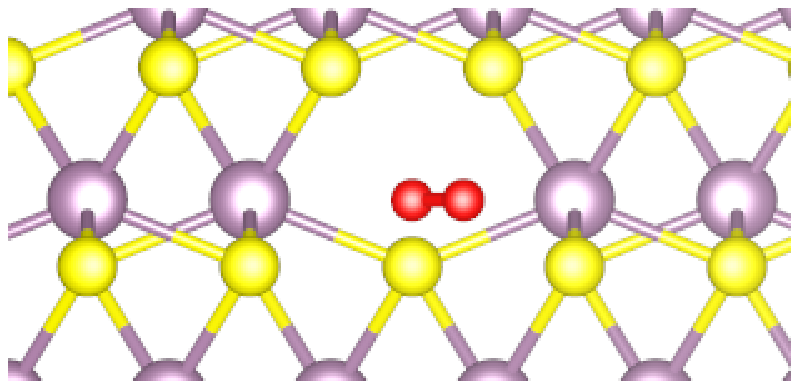}
\hspace{0.5cm}
\includegraphics[width=3cm,  keepaspectratio, clip = true]{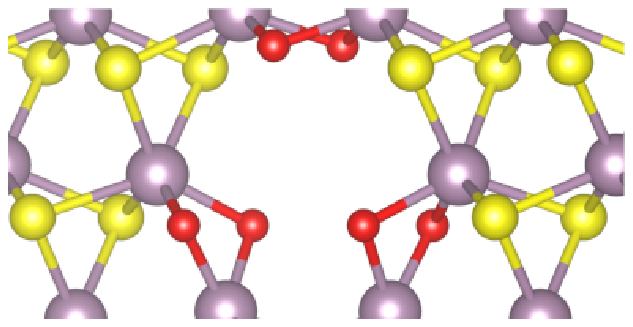}
\hspace{0.5cm}
\includegraphics[width=3cm,  keepaspectratio, clip = true]{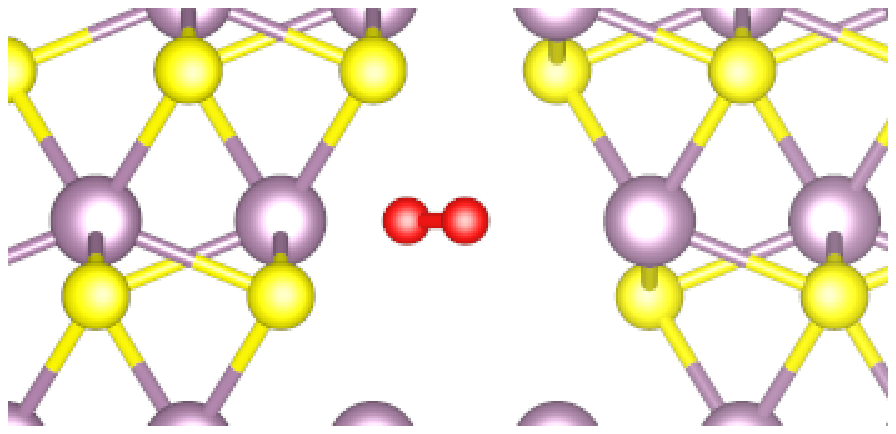}
\caption{\label{fig:oxygengeometry} Top and rotated views of relaxed sructures of oxygen adsorption in  ${\rm MoS_2}$ monolayers. a) and e) pristine, b) and f) V$_{\rm 1Mo-O_2}$, c) and g) V$_{\rm 1Mo6S-3O_2}$ (dissociated) and d) and h) V$_{\rm 1Mo6S-O_2}$ (non-dissociated) defects.}
\end{center}
\end{figure*}

\begin{figure*}[h!]
 \begin{center}
\includegraphics[width=8.0cm,clip = true, keepaspectratio]{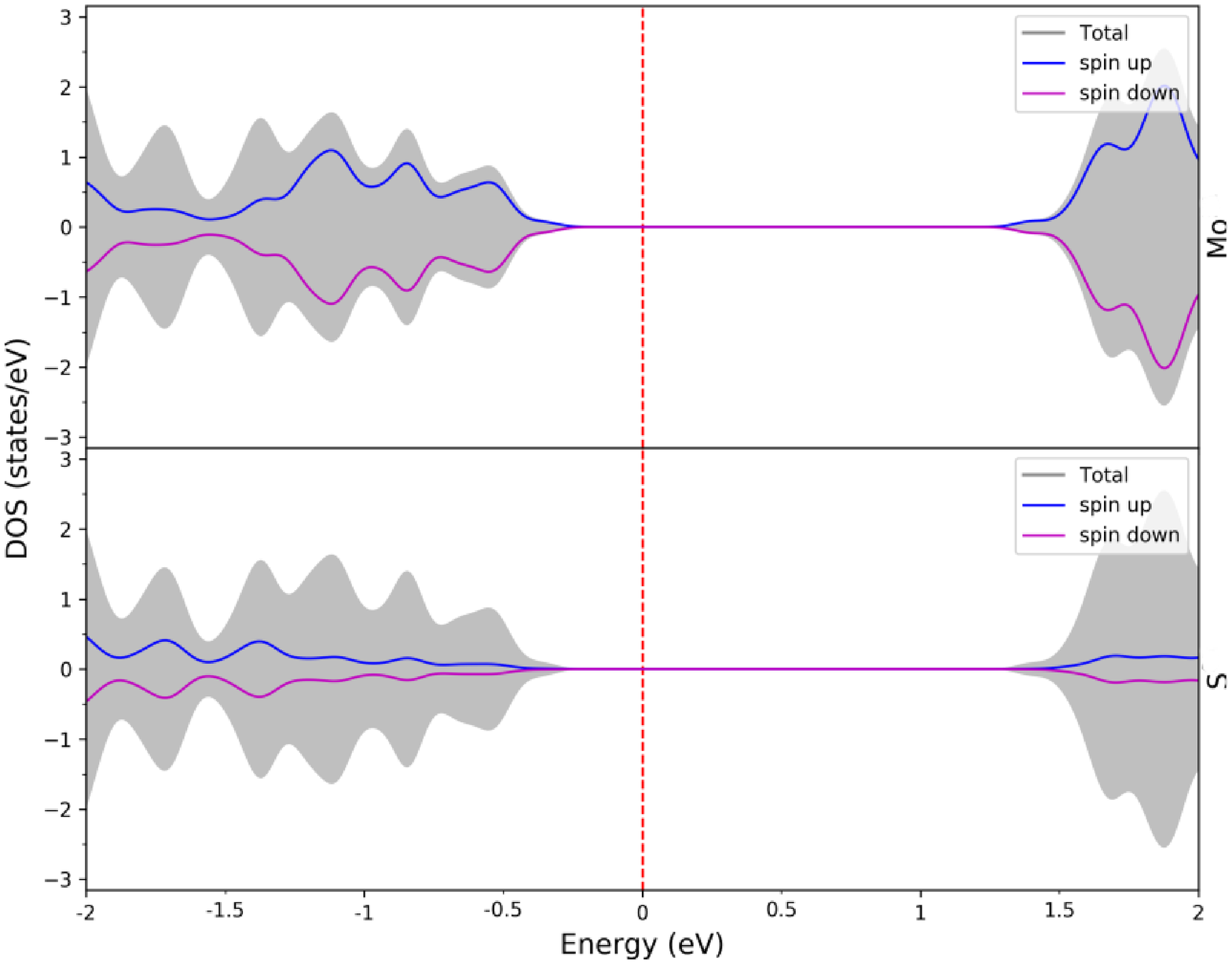}
\includegraphics[width=8.0cm, clip = true, keepaspectratio]{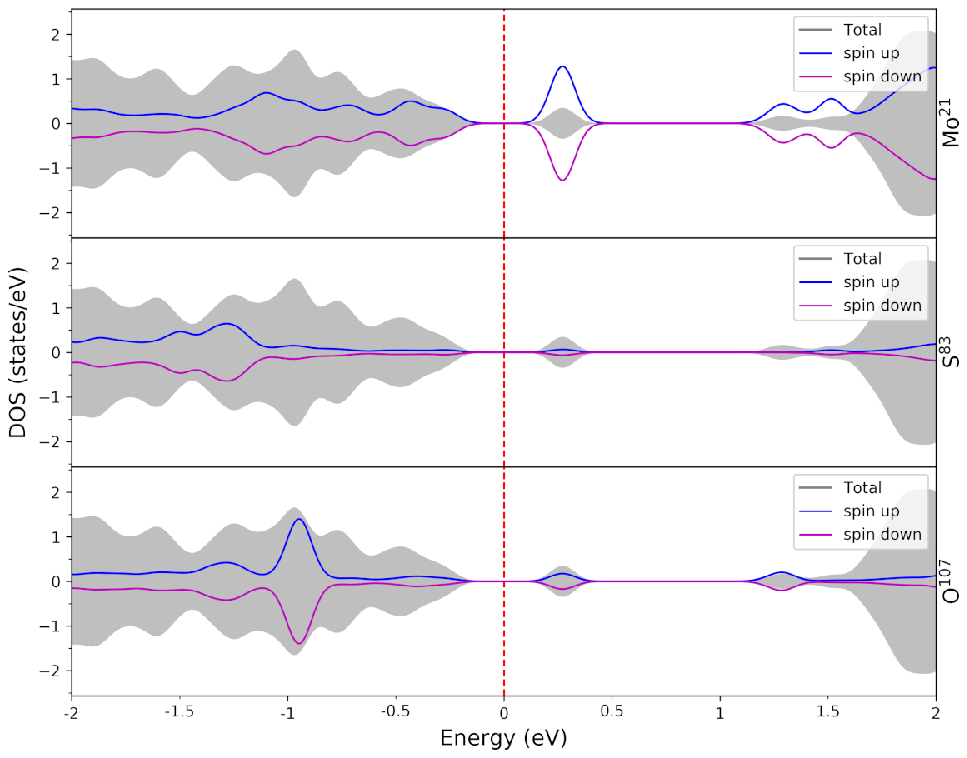}\\
\includegraphics[width=8.0cm,clip = true, keepaspectratio]{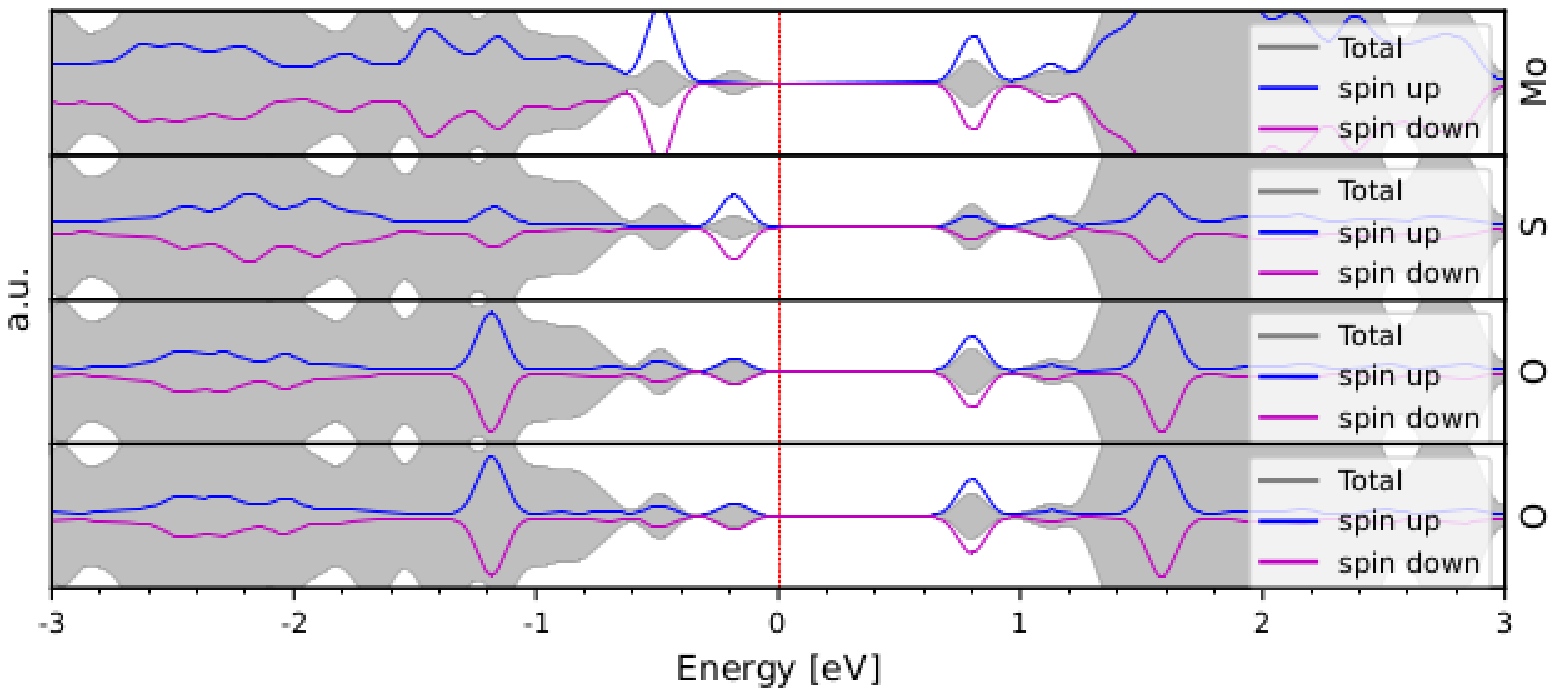}   
\includegraphics[width=8.0cm,clip = true, keepaspectratio]{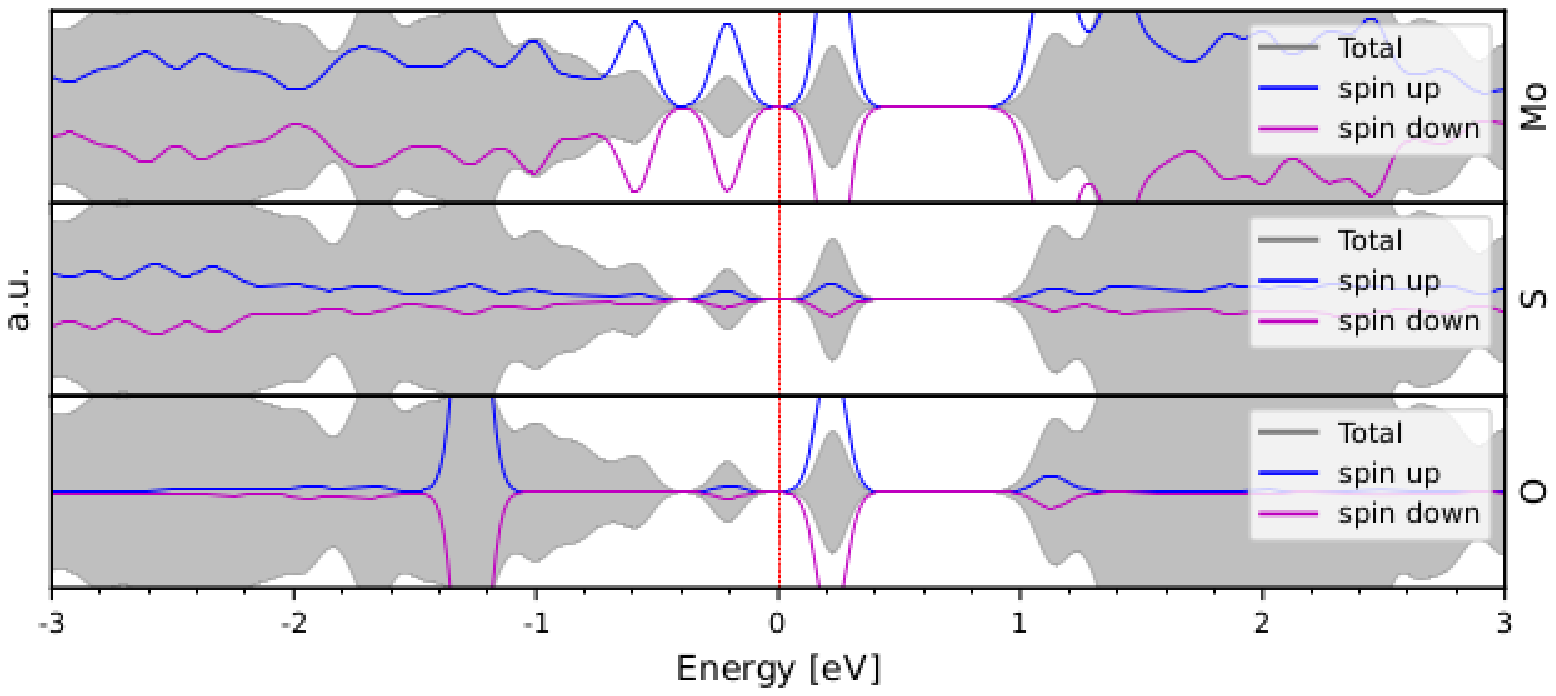}
\caption{\label{fig:dos_o2_defects} DOS and PDOS of oxygen adsorbed in ${\rm MoS_2}$ monolayers: a) pristine and with b) ${\rm V_{1Mo-O_2}}$, c)
  ${\rm V_{1Mo6S-3O_2}}$ (dissociated), d) V$_{\rm 1Mo6S-O_2}$ (non-dissociated) defects.}
\end{center}
\end{figure*}

On the monolayer pristine the oxygen molecule spontaneously
dissociates and the oxygen atoms diffuses further apart to finally
bind on the sulfur atoms. This means that the barrier for O diffusion
should be low, as suggested in Ref.\,\cite{Rao2017}. The binding
energy of the final configuration is -1.13 eV/atom.  In addition we
have investigated the interaction of O$_2$ with other small defects
which we call subnanometer pores: a Mo single vacancy, V$_{\rm Mo-O_2}$, shown in
Fig.\,\ref{fig:oxygengeometry} (b), a Mo single vacancy plus S
hexavacancy, $V_{\rm 1Mo6S-O_2}$, shown in Fig.\,\ref{fig:oxygengeometry}
(c) and d) V$_{\rm 1Mo6S-O_2}$ (non-dissociated). Such structures can
serve as prototypes for nanopores in MoS$_2$.  Additionally, we have
considred a Mo triple vacancy plus a S divacancy, V$_{\rm 3Mo2S-O_2}$, and
a single Mo vacancy plus a S tetravacancy, V$_{\rm 1Mo4S-O_2}$, a Mo
triple vacancy plus a S hexavacancy, V$_{\rm 3Mo6S-O_2}$ (to be published).

Binding energies of oxygen for the relaxed structures are shown in
Table\,\ref{tab:adsorption}. On pristine MoS$_2$, shown in
Fig.\,\ref{fig:oxygengeometry} (a), adsorption of oxygen leads to a
dissociative configuration with binding energy is -0.55\,eV/O atom. Oxygen
spontaneously dissociates with the oxygen atoms diffusing further
apart with a 6.3\,{\AA} being the smallest O-O distance. The O sits
right above the S atom with S-O distance equal to 1.48\,{\AA}.

On V$_{\rm 1Mo-O_2}$ oxygen chemisorbs in a
non-dissociate manner with binding energy of -0.37\,eV/O atom.  
The O-O distance is 2.78\,{\AA} with each oxygen bound to a sulphur atom
in different layers. The relaxed geometry is shown in
Fig.\,\ref{fig:oxygengeometry} (b).  

In a single sulfur vacancy defect, V$_{\rm 1S-O_2}$, shown in Fig.\,\ref{fig:oxygengeometry} (c), the oxygen molecule
dissociates.  Upon dissociation, the first atom goes to a
substitutional sulphur site, while the second oxygen atom moves to
ontop of a sulphur atom. This is in agreement with results reported
previously\,\cite{JPCC2021}.  The binding energy is -0.90 eV/O atom.

On a double sulphur vacancy, V$_{\rm 2S-O_2}$ both oxygen atoms go to
substitutional sulphur positions. The Mo-O bond length is 1.7\,{\AA}
forming a slightly distorted hexagon. One notes that typical Mo-O bond lengths range from 1.69 to 1.73\,{\AA}\,\cite{Hardcastle1990}.  
The O-O distance is our case 2.38\,{\AA}.  The binding energy amounts -3.44\,eV/atom, meaning that the reaction is exothermic.

The V$_{\rm 1Mo6S-O_2}$ interacts with the oxygen molecule via
a perpendicular orientation of the molecule with respect to the MoS$_2$ surface  in its final
configuration. The  oxygen distance to the molybdenum atoms is around
3\,{\AA}.

For other large nanopores such as V$_{\rm 3Mo6S-O_2}$, V$_{\rm 1Mo4S-O_2}$ and V$_{\rm  3Mo2S-O_2}$ (to be published)), the binding energy indicates
physisorption, with the O$_2$ molecule lying in the middle of the pore
with O-O bond distance equals to 1.0\,{\AA}. We may argue here that
there is an interplay between  the number
of oxygen atoms and the number of  dangling bonds in the pore.

The density of states of the above discussed nanopores are show in
Fig.\,\ref{fig:dos_o2_defects}.  V$_{\rm 1S-O_2}$ shows states within
the gap, as seen in Fig.\,\ref{fig:dos_o2_defects}(a). On the other hand, V$_{\rm 2S-O_2}$ does not have states in the gap as shown in Fig.\,\ref{fig:dos_o2_defects}(b).  
Furthermore, V$_{\rm 1Mo-O_2}$ and V$_{\rm 1Mo6S-O_2}$ have states within the gap due to sulphur dangling bonds.  

Recently it has been suggested that the
formation of oxygen at substitutional sites with no band gap states
can be formed in MoS$_2$ layers\,\cite{Small2020}.  A possible
explanation for the ordered oxygen incorporation is that due to the
higher strength of the Mo-O bonds compared to M-S, the substitutional
oxidation of the 2D MoS$_2$ basal plane could be also
thermodynamically favourable\,\cite{Longo2017,Martincova2017}.
One can explain the dissociation considering that SO$_2$ has cohesive
energy of -3.1\,eV, which is larger than the formation enthalpy of
MoS$_2$ which is -2.48\,eV.

\begin{figure*}
\begin{center}
\includegraphics[width=8.0cm, clip= true]{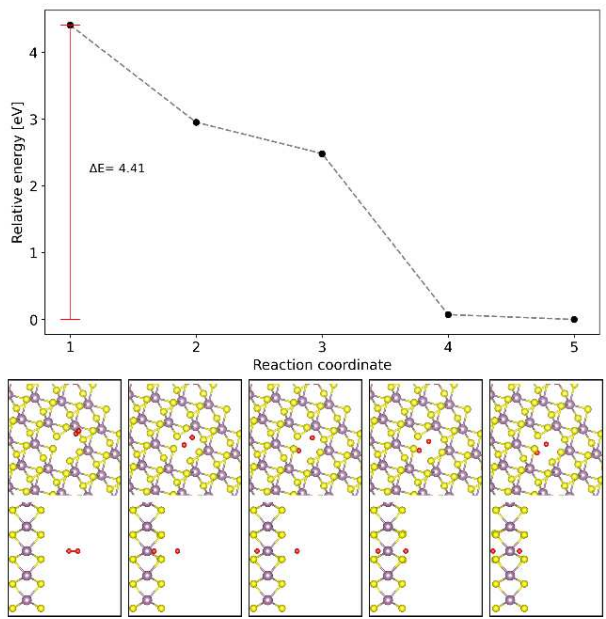}
\caption{\label{fig:barrier_oxygenmol} Diffusion barrier (top panel)  and corresponding paths (lower panel) of an oxygen molecule on a molibdenum vacancy in MoS$_2$  with perpendicular orientation  to the MoS$_2$ basal plane.}
\end{center}
\end{figure*}

Experiments reveal relatively low activation energy values for MoS$_2$
oxidation, ranging from 0.54\,eV to 0.98\,eV\,\cite{Rao2017}. Recent
theoretical calculations\,\cite{JPCC2021} report a barrier of 0.33\,eV in the presence of sulphur defects.
Here we considered the presence of a molibdenum  vacancy. The reaction path
is done by following the molecule along its minimum energy path. We
have used between five and eight intermediate states to calculate the
diffusion.  While oxygen atom diffuses through the pore with a barrier of
2.53\,eV. On the other hand,  an oxygen molecule diffuses barrierless along the path shown
in Fig.\,\ref{fig:barrier_oxygenmol} respectively.

\begin{figure*}
\begin{center}
 \includegraphics[width=4.0cm, clip= true]{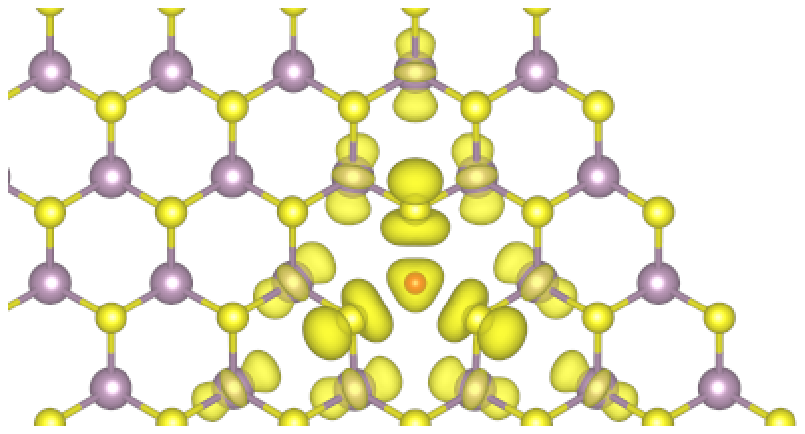}
\hspace{0.5cm}
\includegraphics[width=4.0cm, clip= true]{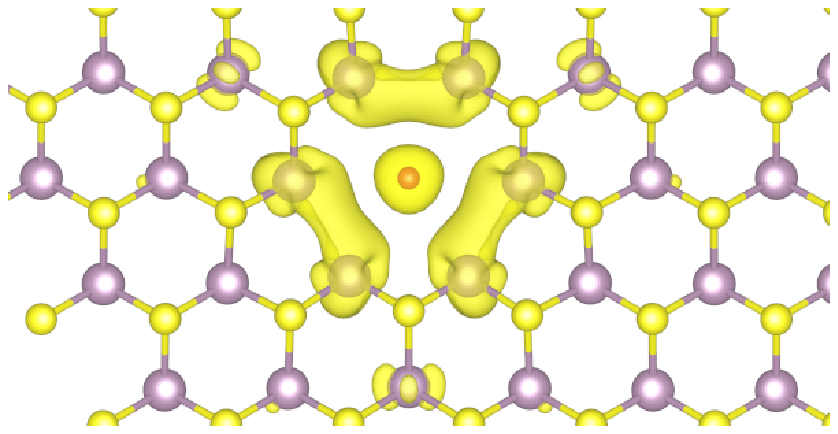}
\hspace{0.5cm} 
\includegraphics[width=4.0cm, clip= true]{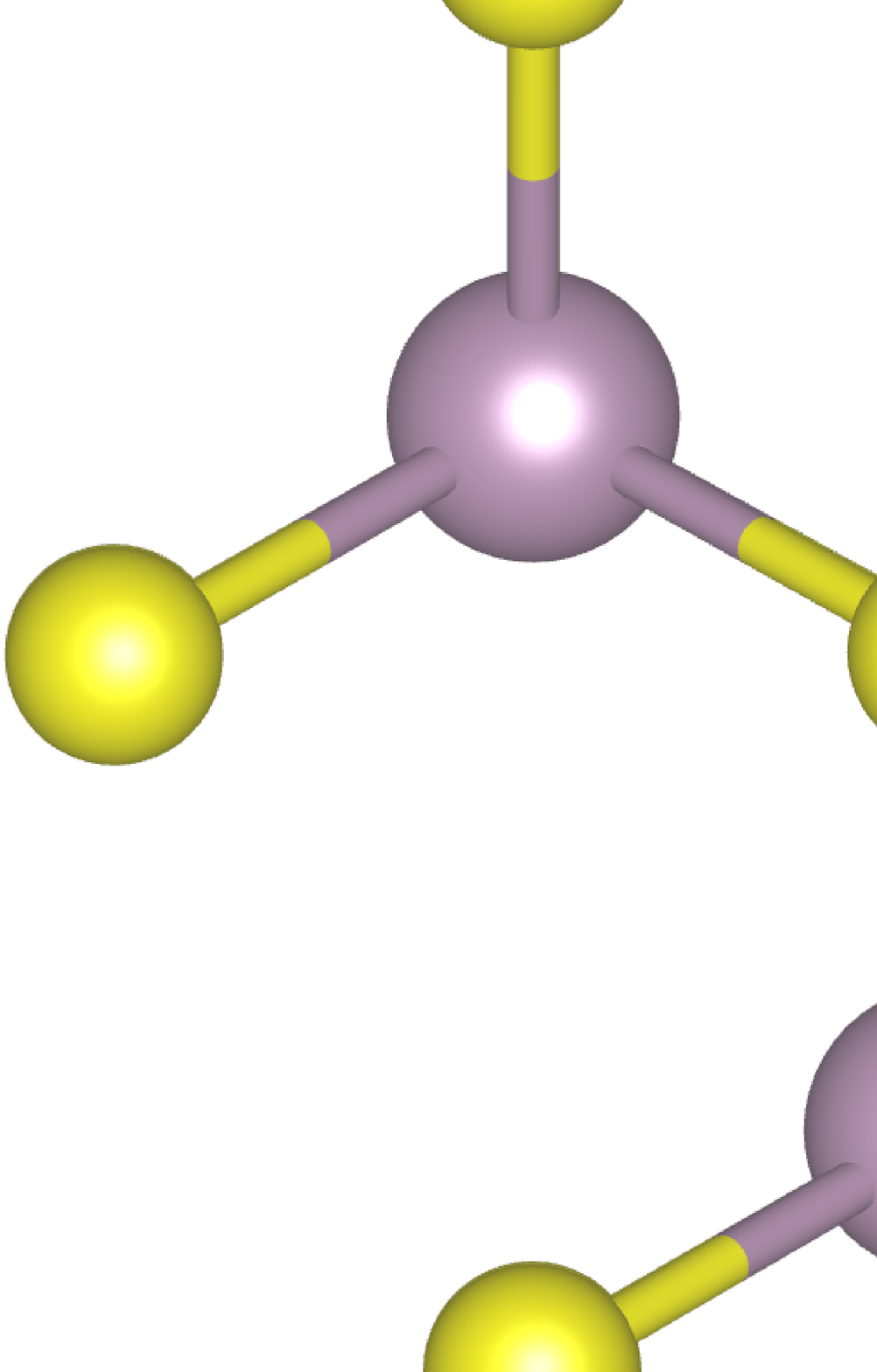}
 \caption{\label{fig:parchg} Projected charge density on LUMO-1 of a) V$_{\rm 1Mo-O_2}$, b) V$_{\rm 1Mo6S-O_2}$ and c) V$_{\rm 3Mo6S-O_2}$. Isosurface value is 0.003 e/bohr$^3$}
\end{center}
\end{figure*}

In order to have further insight on the interaction between the  oxygen molecules and
the nanopores we have calculated the projected partial charge density on the
first level above the Fermi level. These are shown in
Fig.\,\ref{fig:parchg} (a) for V$_{\rm 1Mo-O_2}$ which has the unnocupied orbitals of the pore
oriented towards the molecule.  However, as the preferred orientation
of the molecule in the middle of the pore is perpendicular to the
pore, this could explain the small overlap between HOMO of the
molecule and LUMO of the pore and therefire the relative small berrier for
V$_{\rm 1Mo-O_2}$ where the molecule easily diffuses through the pore.

For the V$_{\rm 1Mo-O_2}$ the molecule diffuses without barrier. Furthermore, the
V$_{\rm 3Mo6S-O_2}$ pore is large enough for the molecule to diffuse
either parallel or perpendicular to the pore. However, if the molecule
is close enough to the edges, it is likely to dissociate. On the other
hand, the V$_{\rm 1Mo6S-O_2}$ pore have its Mo orbitals hybridizing 
forming metallic-like bonds.

\section{Conclusions}

In this work we carried out first-principles
calculations of oxygen on monolayered MoS$_2$  and sub-nanometer MoS$_2$
nanopores. The dissociation and diffusion of oxygen in MoS$_2$ reveal two main features:
the orientation of the molecule with respect to the surface plays a role  on the molecule diffusion barrier;
the reactivity of the pore plays a role on the molecule dissociation.

\section{Acknowledgements}

We acknowledge financial support from Conselho Nacional de Pesquisa e
Desenvolvimento (CNPq) under grants 313081/2017-4 and 305335/2020-0. The calculations have been performed using the
computational facilities of Santos Dumont Supercomputar at LNCC, 
CENAPAD at Unicamp and LaMCAD/UFG.

\bibliographystyle{apsrev}

\end{document}